\documentclass[letterpaper,twocolumn,10pt]{article}
\usepackage{usenix-2020-09}
\usepackage{xspace}
\usepackage{booktabs}
\usepackage{graphicx}
\usepackage{multirow}
\usepackage{amsmath}
\usepackage{tikz}
\usetikzlibrary{positioning,arrows.meta,shapes.misc,fit}
\usepackage{xcolor}
\usepackage{enumitem}
\setlist{nosep,leftmargin=*}

\newcommand{\ADJLAuc}{0.93\xspace}

\newcommand{\ADJLFprHalfT}{11.8\xspace}
\newcommand{\ADJLTpr}{55.3\%\xspace}
\newcommand{\ADJLTprCI}{26.6--75.2}

\newcommand{\ADJLTprHalfT}{87.0\xspace}
\newcommand{\ADJLTprT}{55.3\xspace}
\newcommand{\ADJQAuc}{0.90\xspace}

\newcommand{\ADJQFprHalfT}{6.7\xspace}
\newcommand{\ADJQTpr}{57.4\%\xspace}
\newcommand{\ADJQTprCI}{52.5--62.7}

\newcommand{\ADJQTprHalfT}{77.5\xspace}
\newcommand{\ADJQTprT}{57.4\xspace}
\newcommand{\ADPAAuc}{0.82\xspace}

\newcommand{\ADPAFprHalfT}{20.0\xspace}

\newcommand{\ADPANTpr}{13.2\%\xspace}

\newcommand{\ADPATpr}{0.5\%\xspace}
\newcommand{\ADPATprCI}{0.0--1.2}

\newcommand{\ADPATprHalfT}{58.6\xspace}
\newcommand{\ADPATprT}{0.5\xspace}
\newcommand{\ADPGAuc}{0.83\xspace}

\newcommand{\ADPGFprHalf}{20.0\%\xspace}
\newcommand{\ADPGFprHalfT}{20.0\xspace}

\newcommand{\ADPGNFprHalf}{6.7\%\xspace}

\newcommand{\ADPGNTpr}{3.0\%\xspace}
\newcommand{\ADPGNTprHalf}{74.3\%\xspace}

\newcommand{\ADPGTpr}{2.1\%\xspace}
\newcommand{\ADPGTprCI}{0.0--4.4}
\newcommand{\ADPGTprHalf}{73.1\%\xspace}
\newcommand{\ADPGTprHalfT}{73.1\xspace}
\newcommand{\ADPGTprT}{2.1\xspace}
\newcommand{\ADSJLAuc}{0.93\xspace}

\newcommand{\ADSJLFprHalfT}{11.8\xspace}

\newcommand{\ADSJLTprCI}{32.9--79.9}

\newcommand{\ADSJLTprHalfT}{88.1\xspace}
\newcommand{\ADSJLTprT}{61.9\xspace}
\newcommand{\ADSJQAuc}{0.92\xspace}

\newcommand{\ADSJQFprHalfT}{6.7\xspace}

\newcommand{\ADSJQTprCI}{61.1--70.7}

\newcommand{\ADSJQTprHalfT}{81.5\xspace}
\newcommand{\ADSJQTprT}{64.4\xspace}
\newcommand{\ADSPAAuc}{0.84\xspace}

\newcommand{\ADSPAFprHalfT}{20.0\xspace}

\newcommand{\ADSPATprCI}{0.3--1.5}

\newcommand{\ADSPATprHalfT}{65.7\xspace}
\newcommand{\ADSPATprT}{0.8\xspace}
\newcommand{\ADSPGAuc}{0.84\xspace}

\newcommand{\ADSPGFprHalfT}{20.0\xspace}

\newcommand{\ADSPGTprCI}{3.1--12.4}

\newcommand{\ADSPGTprHalfT}{80.7\xspace}
\newcommand{\ADSPGTprT}{7.9\xspace}
\newcommand{\BIJLAuc}{0.59\xspace}

\newcommand{\BIJLFprHalfT}{4.8\xspace}

\newcommand{\BIJLTprCI}{6.4--11.1}

\newcommand{\BIJLTprHalfT}{15.9\xspace}
\newcommand{\BIJLTprT}{8.3\xspace}
\newcommand{\BIJQAuc}{0.81\xspace}

\newcommand{\BIJQFprHalfT}{1.4\xspace}

\newcommand{\BIJQTprCI}{11.0--18.5}

\newcommand{\BIJQTprHalfT}{16.6\xspace}
\newcommand{\BIJQTprT}{14.9\xspace}
\newcommand{\BIPAAuc}{0.32\xspace}

\newcommand{\BIPAFprHalfT}{12.6\xspace}

\newcommand{\BIPATprCI}{0.4--1.3}

\newcommand{\BIPATprHalfT}{9.0\xspace}
\newcommand{\BIPATprT}{0.7\xspace}
\newcommand{\BIPGAuc}{0.99\xspace}

\newcommand{\BIPGFprHalfT}{0.2\xspace}

\newcommand{\BIPGTpr}{95.1\%\xspace}
\newcommand{\BIPGTprCI}{93.1--96.1}

\newcommand{\BIPGTprHalfT}{92.6\xspace}
\newcommand{\BIPGTprT}{95.1\xspace}
\newcommand{\BISplitPGTestTpr}{95.1\%\xspace}
\newcommand{\BISplitPGTrainTpr}{97.9\%\xspace}
\newcommand{\BestBipiaRankOnAD}{thirteen\xspace}

\newcommand{\CleanPABlock}{69.1\%\xspace}
\newcommand{\CleanPABlockCI}{60--78\%}
\newcommand{\CleanPAFPR}{30.1\%\xspace}

\newcommand{\CleanPGBlock}{59.8\%\xspace}
\newcommand{\CleanPGBlockCI}{51--70\%}
\newcommand{\CleanPGFPR}{29.5\%\xspace}
\newcommand{\DegDeepsetReadme}{100.0\%\xspace}
\newcommand{\DegFmopsReadme}{100.0\%\xspace}
\newcommand{\FmtPAReadmeOrig}{9.8\%\xspace}
\newcommand{\FmtPAReadmePlain}{4.2\%\xspace}
\newcommand{\FmtPGReadmeOrig}{24.6\%\xspace}
\newcommand{\FmtPGReadmePlain}{5.6\%\xspace}

\newcommand{\GJQAct}{61\%\xspace}

\newcommand{\GJQPas}{69\%\xspace}
\newcommand{\GJQRel}{1\%\xspace}

\newcommand{\GPGIrr}{93\%\xspace}

\newcommand{\GPGoIrr}{39\%\xspace}

\newcommand{\GSHAct}{16\%\xspace}
\newcommand{\GSHIrr}{20\%\xspace}

\newcommand{\GSHRel}{87\%\xspace}
\newcommand{\GSHTgt}{88\%\xspace}

\newcommand{\GenPAEmailFPR}{1.2\%\xspace}

\newcommand{\GenPAWebFPR}{0.5\%\xspace}

\newcommand{\MHZAD}{82.2\%\xspace}

\newcommand{\MHZBlock}{0.0\%\xspace}
\newcommand{\MHZClean}{0.0\%\xspace}

\newcommand{\MPGtAD}{69.4\%\xspace}

\newcommand{\MPGtClean}{0.6\%\xspace}

\newcommand{\MPGtsAD}{60.9\%\xspace}

\newcommand{\MPGtsClean}{0.0\%\xspace}

\newcommand{\MPRAD}{72.2\%\xspace}

\newcommand{\MSHAD}{20.1\%\xspace}
\newcommand{\MSHADAuc}{0.86\xspace}

\newcommand{\MSHClean}{10.6\%\xspace}
\newcommand{\MSHGAD}{30.3\%\xspace}

\newcommand{\MSHGClean}{28.6\%\xspace}

\newcommand{\MSHTAD}{9.0\%\xspace}
\newcommand{\MSHTADAuc}{0.91\xspace}

\newcommand{\MSHTClean}{26.0\%\xspace}

\newcommand{\MTSAD}{27.8\%\xspace}

\newcommand{\MWFAD}{73.8\%\xspace}

\newcommand{\MWFBlock}{3.1\%\xspace}
\newcommand{\MWFClean}{0.9\%\xspace}

\newcommand{\NADBenign}{585\xspace}
\newcommand{\NADInj}{432\xspace}
\newcommand{\NADSBenign}{975\xspace}
\newcommand{\NADSInj}{720\xspace}
\newcommand{\NBIBenign}{1{,}200\xspace}
\newcommand{\NBIInj}{2{,}435\xspace}
\newcommand{\NCleanOutputs}{339\xspace}
\newcommand{\NCleanTasks}{97\xspace}
\newcommand{\NDetectors}{fifteen\xspace}
\newcommand{\NGeneral}{7{,}504\xspace}
\newcommand{\NIASeen}{53 of 62\xspace}
\newcommand{\NIAUnseen}{9\xspace}
\newcommand{\NTBBenign}{1{,}533\xspace}
\newcommand{\NTBInj}{584\xspace}

\newcommand{\NTBTasks}{675\xspace}

\newcommand{\OvCleanCode}{22\%\xspace}
\newcommand{\OvCleanEmail}{31\%\xspace}
\newcommand{\OvCleanTable}{0.3\%\xspace}
\newcommand{\OvCodeTest}{0 of 50\xspace}
\newcommand{\OvCodeTrain}{6 of 50\xspace}
\newcommand{\OvTextTest}{0 of 75\xspace}
\newcommand{\OvTextTrain}{65 of 75\xspace}
\newcommand{\PGBipiaChars}{715\xspace}
\newcommand{\PGBipiaRows}{1{,}116\xspace}
\newcommand{\PGIAChars}{178\xspace}
\newcommand{\PGIARows}{111\xspace}

\newcommand{\RankTau}{0.01\xspace}

\newcommand{\RankTauP}{1.00\xspace}

\newcommand{\TBFPRHZ}{0.7\%\xspace}

\newcommand{\TBFPRPA}{66.9\%\xspace}

\newcommand{\TBFPRPGt}{0.0\%\xspace}

\newcommand{\TBFPRPR}{46.2\%\xspace}

\newcommand{\TBFPRSH}{4.4\%\xspace}
\newcommand{\TBFPRSHG}{85.6\%\xspace}
\newcommand{\TBFPRSHT}{69.7\%\xspace}

\newcommand{\TBFPRWF}{0.0\%\xspace}

\newcommand{\TBSeenPG}{52.1\%\xspace}

\newcommand{\TBTprHZ}{100.0\%\xspace}

\newcommand{\TBTprJLl}{78.3\%\xspace}
\newcommand{\TBTprJQ}{54.3\%\xspace}

\newcommand{\TBTprPG}{52.7\%\xspace}

\newcommand{\TBTprPGt}{57.9\%\xspace}
\newcommand{\TBTprPGts}{60.8\%\xspace}
\newcommand{\TBTprPR}{15.2\%\xspace}

\newcommand{\TBTprSH}{95.2\%\xspace}
\newcommand{\TBTprSHG}{92.3\%\xspace}

\newcommand{\TBTprWF}{90.8\%\xspace}

\newcommand{\TBUnseenPG}{55.8\%\xspace}

\newcommand{\TauADTB}{0.28\xspace}

\newcommand{\TauADTBP}{0.17\xspace}
\newcommand{\TauBITB}{0.31\xspace}
\newcommand{\TauBITBP}{0.11\xspace}
\newcommand{\TauFPRADTB}{0.67\xspace}
\newcommand{\TauFPRADTBP}{0.001\xspace}

\newcommand{\finding}[2]{\par\smallskip\noindent\textbf{Finding #1.} \emph{#2}\par\smallskip}
\newcommand{\RQ}[1]{\textbf{RQ#1}}

\begin{document}
\date{}
\title{\Large \bf Passing the Test You Trained On:\\Re-evaluating Prompt-Injection Detectors for LLM Agents}
\author{
{\rm Zhuowen Liu}\\
Cybersecurity Lab\\
Japan Advanced Institute of Science and Technology (JAIST)\\
Nomi, Ishikawa, Japan\\
\texttt{s2410431@jaist.ac.jp}
}
\maketitle

\begin{abstract}
LLM agents increasingly screen tool outputs with small prompt-injection detectors, and teams choose among detectors by their scores on public benchmarks. We ask whether those scores predict how a detector behaves inside an agent. We replay the ground-truth tool calls of two agent benchmarks, AgentDojo and $\tau$-bench, without an LLM to obtain tool outputs that are benign by construction, label injected outputs by differential replay, and evaluate fifteen detectors, including Meta's Prompt Guard~2, and two task-aware LLM judges on these outputs and on the BIPIA benchmark. Detection rankings transfer poorly between benchmarks: the best detector on BIPIA catches 2\% of AgentDojo injections at a 1\% false-positive rate, and a detector that catches 72\% of AgentDojo injections catches 15\% on $\tau$-bench. False-positive rates on tool outputs, which range from none to over 90\%, do transfer between the two agent benchmarks. Where training data is public, the form of the training inputs explains the results. The BIPIA leader was trained on full BIPIA inputs, but having seen InjecAgent's attack strings as short prompts does not help it find them inside tool outputs; the best detector on both agent benchmarks shares no data with any benchmark and was trained on agent-style inputs. Evaluations meant to inform deployment should use the agent's own tool outputs, report detection at a low false-positive rate, and audit what the detector was trained on.
\end{abstract}

\section{Introduction}
\label{sec:intro}

LLM agents read email, calendars, web pages, and transaction histories on a user's behalf, and every one of those sources can carry text written by someone else. Indirect prompt injection turns that text into instructions~\cite{greshake2023}. The cheapest defense to deploy is a detector: a small classifier that screens every tool output before the agent sees it and drops what looks like an injection~\cite{protectai,piguard,llamafirewall}. It needs no change to the agent or the model, which is why guardrail toolkits ship detectors and practitioners compare them.

The threat is simple to state. An attacker who can write a calendar event description, a product review, or a table cell controls part of what a tool returns; if the detector misses that text, the agent may act on it. If the detector is too eager instead, it drops ordinary tool outputs, and the agent cannot finish the user's task.

How would a team decide which detector to deploy? Detector papers and third-party comparisons report accuracy on prompt collections and injection benchmarks; a 2025 evaluation of nine open guardrails, for example, concluded from BIPIA's email and table tasks that PIGuard detects indirect injection effectively~\cite{mozillaguardrails}. Two things in this practice are unchecked. First, benchmark text is not what an agent's detector screens. Agents see serialized tool outputs (records, lists, confirmations), and false alarms on those translate into failed tasks, not into a percentage on a balanced test set. Second, nobody checks whether the benchmark overlaps with the detector's training data. Prior work showed that same-source splits inflate scores for classifiers the authors train themselves~\cite{benchmarkslie} and argued for low-FPR evaluation~\cite{promptshield}; neither audits released detectors or evaluates them where an agent would use them.

We evaluate fifteen detectors, from 2023 DeBERTa classifiers through Meta's Prompt Guard~2 to 2026 models built for agent inputs, together with two task-aware LLM judges, on the tool outputs of two agent benchmarks, AgentDojo~\cite{agentdojo} and $\tau$-bench~\cite{taubench}, and on BIPIA~\cite{bipia}. Our central observation is that a detector works on inputs whose form resembles what it was trained on, and that benchmark scores mostly measure this resemblance rather than a general ability to detect injections.

Making this measurement requires three things that existing benchmarks do not provide. First, benign agent tool outputs with certain ground truth: we replay the benchmarks' ground-truth tool calls without any LLM, which yields tool outputs that are benign by construction. Second, reliable labels for injected outputs: substring matching fails because AgentDojo re-wraps injected text when it serializes outputs, so we label by differential replay against the clean environment. Third, a way to connect performance to training data: we audit the published training data of two detectors against all three benchmarks and against InjecAgent's attacks~\cite{injecagent}, and we include a judge whose training data predates AgentDojo.

The results reverse the conclusions a benchmark-based comparison would draw. Detection rankings transfer poorly: between BIPIA and AgentDojo Kendall's $\tau$ is \RankTau, and even between the two agent benchmarks it is \TauADTB. PIGuard catches \BIPGTpr of BIPIA injections at 1\% FPR and \ADPGTpr of AgentDojo injections; Prismor catches \MPRAD of AgentDojo injections and \TBTprPR of $\tau$-bench's. False-positive rates on tool outputs, by contrast, range from none to over 90\% and are consistent across the two agent benchmarks ($\tau=\TauFPRADTB$). The training data explains the pattern. PIGuard was trained on \PGBipiaRows full BIPIA inputs; it also saw \NIASeen InjecAgent attacks, but only as short prompts, and inside $\tau$-bench tool outputs it detects those attacks no better than ones it never saw (\TBSeenPG versus \TBUnseenPG). Horizon-Labs shares no data with any benchmark, was trained on agent-style inputs, and leads both agent benchmarks (\MHZAD and \TBTprHZ) with almost no false alarms. Task-aware judges reach 54--78\% on the agent benchmarks, below the best dedicated detectors, and every approach misses whole classes of injections outside its training distribution.

We make the following contributions:
\begin{itemize}
\item A methodology for evaluating injection detectors where agents use them: clean replay of ground-truth tool calls for benign outputs, differential replay for injected ones, and task-level block rates (\S\ref{sec:method}).
\item A measurement of fifteen detectors and two judges on two agent benchmarks and BIPIA, showing that false-positive behavior on tool outputs is stable across agent benchmarks while detection rankings transfer poorly (\S\ref{sec:rq1}, \S\ref{sec:rq2}).
\item Training-data audits and a controlled seen/unseen comparison showing that the form of the training inputs, not overlap in attack strings, decides where a detector works (\S\ref{sec:rq3}).
\item An analysis of what task awareness, attack goals, and input format change, with the finding that task-irrelevant injections evade every approach outside its training distribution (\S\ref{sec:rq4}, \S\ref{sec:rq5}).
\end{itemize}
The code that rebuilds every evaluation set from the public benchmarks and reproduces every number in this paper is available at \url{https://github.com/lzwhehe/benign-instruction-bench}.

\section{Background and Motivating Example}
\label{sec:background}

\paragraph{Indirect prompt injection.}
An LLM agent cannot tell instructions from data by position alone: text that a tool returns can redirect the model as easily as text the user typed~\cite{greshake2023,liu2024formalizing}. Defenses fall into three families. Model-level defenses train the model to ignore instructions in the data region~\cite{struq,secalign}. System-level defenses keep untrusted data away from the component that plans actions~\cite{camel}. Detection-based defenses run a separate classifier on untrusted text and drop what it flags~\cite{protectai,piguard,promptshield,datasentinel}. Detectors are the easiest to adopt, because they need no change to the agent or the model, and they are the subject of this paper.

\paragraph{How detectors are evaluated.}
Detector papers report accuracy on collections of prompts and injected documents, often including BIPIA~\cite{bipia} and the over-defense set NotInject~\cite{piguard}. Practitioners compare released detectors the same way: a 2025 evaluation of nine open guardrails used BIPIA's email and table tasks and concluded that PIGuard detects indirect injection effectively~\cite{mozillaguardrails}. Two questions are left open by this practice. The text in these collections is not what an agent's detector sees, which is serialized tool output. And nothing checks whether the benchmark overlaps with what the detector was trained on.

\begin{figure}[t]
\centering
\resizebox{\columnwidth}{!}{%
\begin{tikzpicture}[
  box/.style={draw, rounded corners=2pt, align=left, font=\scriptsize, inner sep=3pt, text width=#1},
  box/.default=3.0cm,
  arr/.style={-{Stealth[length=4pt]}, thick, draw=black!60},
  lab/.style={font=\scriptsize\bfseries}]
\node[box=3.2cm, fill=blue!5] (u) {\textbf{User:} ``Please update my account password and then pay the bill in my files.''};
\node[box=2.2cm, right=0.5cm of u, fill=gray!8, align=center] (a) {\textbf{Agent}\\(LLM + tools)};
\node[box=3.6cm, below=0.55cm of u, fill=green!6, xshift=0.2cm] (o1) {\texttt{update\_password} returns\\\texttt{\{'message': 'Password updated.'\}}};
\node[box=3.6cm, below=0.25cm of o1, fill=red!6] (o2) {\texttt{search\_calendar\_events} returns an event whose description contains text planted by an attacker.};
\node[box=2.2cm, right=0.5cm of o1, yshift=-0.55cm, fill=orange!10, align=center] (d) {\textbf{Detector}\\screens each\\tool output};
\node[box=2.6cm, right=0.45cm of d, yshift=0.55cm, fill=green!3] (r1) {\textbf{Blocked.} PIGuard score 1.00: a benign output stops the task.};
\node[box=2.6cm, right=0.45cm of d, yshift=-0.55cm, fill=red!3] (r2) {\textbf{Passed.} At a 1\% FPR threshold, both detectors let almost every such output through.};
\draw[arr] (u) -- (a);
\draw[arr] (a.south) |- ([yshift=0.15cm]o1.east);
\draw[arr] (o1.east) -- (d);
\draw[arr] (o2.east) -- (d);
\draw[arr] (d.east) -- (r1.west);
\draw[arr] (d.east) -- (r2.west);
\end{tikzpicture}}
\caption{Running example. A detector between the tools and the agent blocks a harmless confirmation message (top) and, once its threshold is set to keep false alarms rare, passes almost every injected output (bottom). Both outputs come from AgentDojo; the scores are measured (\S\ref{sec:rq1}, \S\ref{sec:rq2}).}
\label{fig:example}
\end{figure}
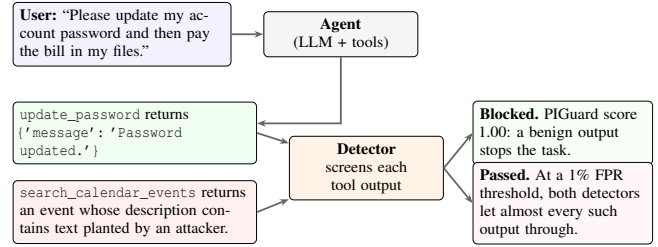

\paragraph{Running example.}
Figure~\ref{fig:example} shows an AgentDojo banking task. The user asks the agent to update a password and pay a bill. The password tool returns \texttt{\{'message': 'Password updated.'\}}, and PIGuard scores this output 1.00, so a deployment that drops flagged outputs fails a task in which nothing malicious happened. In a workspace task, a calendar event carries a description that an attacker wrote. At their default thresholds the detectors catch many such outputs, but they pay for it with false alarms on a fifth of all benign tool outputs. Once the threshold is moved to keep false alarms at 1\%, almost every injected output passes. Yet on BIPIA, the same PIGuard model catches nearly every injection at the same false-alarm rate. Detectors built for agent inputs show the mirror image: no false alarm on the password confirmation, most AgentDojo injections caught, and most BIPIA injections missed. The rest of this paper explains the pattern.

\paragraph{Why not just pick a better threshold, or a bigger model?}
Thresholds trade one failure for the other; the question is whether any threshold gives both low FPR and useful detection, which is what TPR at 1\% FPR measures. A bigger general-purpose model can serve as a judge, and we test two such judges; but a judge's verdict depends on what it is asked, and \S\ref{sec:rq4} shows that the question we ask covers some attack goals and not others.

\section{Deployment Setting and Threat Model}
\label{sec:threat}

\paragraph{System model.}
A user gives an LLM agent a task in natural language. The agent calls tools (email search, calendar lookup, transaction history, web pages, file reads) and appends each tool output to its context before deciding the next step. A \emph{detector} screens every tool output before it enters the context. If the detector flags an output, the deployment either drops the output or stops the task and asks a human. Either way, a flagged benign output costs the user the task, so we count a task as \emph{blocked} if any of its tool outputs is flagged. This is the configuration of detector-based defenses in agent frameworks, including the detector baseline that AgentDojo ships~\cite{agentdojo} and the scanner layer of LlamaFirewall~\cite{llamafirewall}.

\paragraph{Attacker capabilities and goals.}
The attacker controls part of the data some tool returns (\textbf{A1}): a calendar event description, an email body, a review on a booking site, a table cell, or an answer on a Q\&A site. The attacker cannot change the user's request, the agent, or the detector. We distinguish attacks by goal, following BIPIA's taxonomy~\cite{bipia}: \emph{action} goals make the agent do something consequential (send money, forward a file, run code); \emph{task-irrelevant} goals make it perform an unrelated task; \emph{task-relevant} goals change how it answers the user's own task (reverse the text, answer in Base64).

\paragraph{Security goals of the detector.}
\textbf{G1}: benign tool outputs pass, so that tasks are not blocked. \textbf{G2}: injected tool outputs are flagged at an operating point that satisfies G1.

\paragraph{Scope.}
We measure detectors against the static attacks that public benchmarks ship. An adaptive attacker who targets a specific detector would lower detection rates further~\cite{attackermovessecond}, so our detection numbers are upper bounds. We do not propose a new defense; the task-aware judges in \S\ref{sec:method} serve as a diagnostic that separates what the benchmark measures from what the detector learned.

\section{Methodology}
\label{sec:method}

\subsection{Detectors and Judges}

\paragraph{Detectors.}
We evaluate \NDetectors detectors (Table~\ref{tab:many}). We started from the most downloaded text-classification models on Hugging Face that are labeled as prompt-injection detectors, kept those that output an injection decision, and added the agent-oriented detectors released in 2026. Six are gated behind a license agreement, which we accepted: Meta's Prompt Guard~1 and Prompt Guard~2 (86M and 22M)~\cite{llamafirewall}, ProtectAI's small v2 model, and two Sentinel models. The set spans three generations: DeBERTa-v3 classifiers from 2023--2025 (ProtectAI~\cite{protectai}, JasperLS, TestSavant, PIGuard~\cite{piguard}, Prompt Guard), ModernBERT and Qwen3-based classifiers from 2025--2026 (the two Sentinels, and Sheltron, Wolf Defender, and Horizon-Labs, whose cards target agent inputs), and a fine-tuned 1.5B generative classifier (Prismor). We run each detector as its model card documents, including its context length (512 to 2{,}048 tokens), its segmentation of long inputs, its input template, and, for Sheltron, its role and source tokens; for Prompt Guard~1 we use Meta's indirect-injection score, the sum of its injection and jailbreak probabilities. Appendix~\ref{app:configs} lists the settings. Two detectors need care. InjecGuard returns the same scores as PIGuard on every input we scored, so we report it once. Sheltron's encoder also accepts a trusted ``user goal'' segment; its card does not document this use, but we report it as a task-aware variant because it is the only detector that can take the user's request as input. Two further detectors, \texttt{deepset/deberta-v3-base-injection} and \texttt{fmops/distilbert-prompt-injection}, flag every benign README chunk and we omit them (Appendix~\ref{app:excluded}). One gated detector (CodeIntegrity) did not grant access.

\paragraph{Task-aware judges.}
To test whether information the detectors lack would change the outcome, we prompt an instruction-tuned model with the user's request and one tool output and ask whether the output tries to make the assistant do something the user did not ask for (prompt in Appendix~\ref{app:prompt}). The score is the probability of ``Yes'' relative to ``No'' at the first generated token, summed over token variants. We use Qwen2.5-7B-Instruct~\cite{qwen25} and Llama-3.1-8B-Instruct~\cite{llama3}, both quantized to 4 bits (NF4). The Llama model's training data ends in December 2023, six months before AgentDojo was released, so it cannot have seen AgentDojo's tasks or attack templates. The 4-bit Qwen judge matches the same model served in bf16 to within 0.02 AUROC on AgentDojo.

\subsection{Data}

Table~\ref{tab:data} lists the evaluation sets: two agent benchmarks, one injection benchmark, and general benign text.

\paragraph{General benign text.}
We sample \NGeneral chunks of 80--220 words, about 2{,}500 each from GitHub READMEs~\cite{githubreadmes}, business emails (AESLC~\cite{aeslc}), and educational web pages (FineWeb-Edu~\cite{fineweb}). Every chunk is benign.

\paragraph{Clean agent tool outputs (AD-Clean).}
AgentDojo v1.2 ships, for every user task, the sequence of tool calls that solves it~\cite{agentdojo}. We execute these calls in the default environment, without any injection and without an LLM, and record each tool output exactly as AgentDojo formats it for the agent. This yields \NCleanOutputs outputs from \NCleanTasks tasks in four suites (workspace, travel, banking, slack). Every output is benign by construction.

\paragraph{Matched agent tool outputs (AD-Matched, AD-Shift).}
We place one of AgentDojo's built-in attacks into every injection point of a task's environment and replay the same ground-truth calls. An output is labeled \emph{injected} if it differs from the output of the same call in the clean environment, and \emph{benign} otherwise. Substring matching is not sufficient: AgentDojo serializes outputs as YAML, which re-wraps multi-line injected text. AD-Matched uses three attack templates (\texttt{important\_instructions}, \texttt{ignore\_previous}, \texttt{direct}; \NADBenign benign and \NADInj injected outputs). AD-Shift uses five others (\texttt{tool\_knowledge}, \texttt{injecagent}, \texttt{system\_message}, and two variants of \texttt{important\_instructions}; \NADSBenign and \NADSInj). All AgentDojo injection goals are actions.

\paragraph{A second agent benchmark ($\tau$-bench).}
$\tau$-bench~\cite{taubench} simulates customer-service agents for a retailer and an airline over JSON databases, and ships ground-truth tool calls for each task. Its authors, domains, tools, and output format (JSON rather than YAML) all differ from AgentDojo's. Replaying the ground-truth calls of all \NTBTasks retail and airline tasks without an LLM gives \NTBBenign tool outputs that are benign by construction (TB-Clean). $\tau$-bench has no injection points, so we add one where a third party plausibly controls text: the catalog name of a product, which a marketplace seller writes. For each retail task that looks up a product, we append an attack instruction from InjecAgent~\cite{injecagent} (MIT license; its base and enhanced settings, eight attacks per task, all 62 attacks used) to the names of the products the task looks up, replay the same calls, and label by differential replay. Orders keep the product name recorded at purchase, as in a real store. This yields \NTBInj injected outputs, which we evaluate against TB-Clean as negatives (TB-Matched). InjecAgent was written by yet another group, so its attacks share neither authors nor templates with AgentDojo or BIPIA.

\paragraph{BIPIA.}
BIPIA~\cite{bipia} provides clean external content for three tasks we use (email question answering, table question answering, and code repair with a Stack Overflow answer) and attack strings in three text groups and two code groups. Each clean context is a benign sample; we insert four randomly drawn attack strings at the start, middle, or end, following BIPIA's insertion rules, to obtain injected samples. BIPIA splits its attack strings into a training and a test half. Our main results use only test-half attacks (\NBIBenign benign, \NBIInj injected); \S\ref{sec:rq3} uses both halves.

\begin{table}[t]
\centering\small
\caption{Evaluation sets. Benign / injected counts. AD = AgentDojo v1.2; TB = $\tau$-bench. The benign side of TB-Matched is TB-Clean.}
\label{tab:data}
\begin{tabular}{@{}lrrl@{}}
\toprule
Set & Benign & Injected & Content \\
\midrule
General text & \NGeneral & -- & README, email, web \\
AD-Clean & \NCleanOutputs & -- & tool outputs, no injection \\
AD-Matched & \NADBenign & \NADInj & tool outputs, 3 attacks \\
AD-Shift & \NADSBenign & \NADSInj & tool outputs, 5 other attacks \\
TB-Matched & \NTBBenign & \NTBInj & $\tau$-bench tool outputs (JSON) \\
BIPIA & \NBIBenign & \NBIInj & email, table, code \\
\bottomrule
\end{tabular}
\end{table}

\subsection{Metrics}

We report the false-positive rate (FPR) and true-positive rate (TPR) at each detector's default threshold of 0.5, AUROC, and the TPR at 1\% FPR. A deployment that screens every tool output sees benign outputs far more often than injected ones, so the low-FPR operating point is the one that matters, as in membership inference~\cite{lira} and deployable injection detection~\cite{promptshield}. We pick the 1\%-FPR threshold on the same benign set we evaluate, which favors the detectors; a deployment would have to pick it in advance. For AD-Clean we also report the share of tasks with at least one flagged output. Intervals are 95\% percentile bootstrap intervals over 1{,}000 resamples, stratified by label (by task for task-level rates).

\subsection{Training-Data Audit}

Two detectors publish enough to audit. PIGuard's authors release its training split. We normalize whitespace and case and search it for (i) the first 80 characters of every BIPIA attack string and (ii) the first 120 characters of every clean BIPIA context. Horizon-Labs releases the script that assembles its training set from public datasets; we download the five sources that contain agent or tool-output injections and search them for BIPIA's attack strings and for markers of AgentDojo's attack templates (for example the opening sentence of \texttt{important\_instructions}). We also search both for InjecAgent's 62 attack instructions. Both searches find only verbatim copies and are lower bounds on overlap. The other detectors name their sources at most, so we report what their model cards state.

\section{Results}
\label{sec:results}

We answer five questions.
\begin{itemize}
\item \RQ{1}: How often do detectors flag benign agent tool outputs, and how many agent tasks does that stop?
\item \RQ{2}: How many injections do detectors catch at 1\% FPR, and does a detector's rank on one benchmark predict its rank on another?
\item \RQ{3}: Does what a detector was trained on explain where it works?
\item \RQ{4}: Does knowing the user's request help, and which attack goals does each approach miss?
\item \RQ{5}: Do input format and metadata change false positives?
\end{itemize}

\begin{table*}[t]
\centering\small
\caption{All detectors on all sets. General: README, email, and web text. FPR at each detector's default threshold on benign tool outputs (AD-Clean, TB-Clean); ``tasks'' is the share of tasks with at least one flagged benign output. TPR at the threshold that gives 1\% FPR on the same set. BIPIA uses test-half attacks. ``Agent data'': the model card states training on agent or tool-output inputs; ``eval.'' means the card reports evaluation on AgentDojo-derived data. $^\dagger$Undocumented use of Sheltron's user-goal segment. Bootstrap intervals for four methods are in Appendix Table~\ref{tab:main}.}
\label{tab:many}
\resizebox{\textwidth}{!}{
\begin{tabular}{@{}lccc|c|cc|cc|ccc|c@{}}
\toprule
 & & & & General & \multicolumn{4}{c|}{AgentDojo} & \multicolumn{3}{c|}{$\tau$-bench} & BIPIA \\
 & & & Agent & FPR & FPR & Tasks & \multicolumn{2}{c|}{TPR @ 1\% FPR} & FPR & Tasks & TPR & TPR \\
Detector & Year & Size & data & (\%) & (\%) & (\%) & Matched & Shift & (\%) & (\%) & @ 1\% & @ 1\% \\
\midrule
JasperLS & 2023 & 184M & no & 99.9 & 78.5 & 97.9 & 23.4 & 22.1 & 93.1 & 99.6 & 18.7 & 4.9 \\
ProtectAI v1 & 2023 & 184M & no & 0.3 & 3.5 & 12.4 & 0.7 & 7.5 & 5.1 & 6.8 & 23.3 & 0.4 \\
ProtectAI v2 & 2024 & 184M & no & 3.8 & 30.1 & 69.1 & 0.5 & 0.8 & 66.9 & 80.3 & 10.1 & 0.7 \\
ProtectAI v2-small & 2024 & 142M & no & 1.6 & 12.1 & 34.0 & 7.6 & 3.1 & 1.4 & 2.8 & 30.1 & 1.4 \\
TestSavant-L & 2024 & 184M & no & 3.8 & 3.2 & 10.3 & 27.8 & 24.7 & 1.8 & 3.4 & 24.5 & 16.3 \\
Prompt Guard 1 & 2024 & 279M & no & 5.8 & 31.9 & 53.6 & 9.5 & 1.5 & 4.8 & 9.3 & 2.9 & 36.0 \\
PIGuard & 2025 & 184M & no & 9.4 & 29.5 & 59.8 & 2.1 & 7.9 & 11.0 & 20.9 & 52.7 & 95.1 \\
Prompt Guard 2 (22M) & 2025 & 71M & no & 0.3 & 0.0 & 0.0 & 60.9 & 63.6 & 0.0 & 0.0 & 60.8 & 31.7 \\
Prompt Guard 2 (86M) & 2025 & 279M & no & 0.1 & 0.6 & 2.1 & 69.4 & 78.8 & 0.0 & 0.0 & 57.9 & 10.4 \\
Qualifire Sentinel & 2025 & 396M & no & 18.2 & 0.9 & 3.1 & 61.1 & 66.4 & 0.0 & 0.0 & 21.9 & 4.6 \\
Sentinel v2 & 2025 & 596M & no & 17.2 & 2.1 & 7.2 & 20.8 & 16.5 & 0.5 & 1.2 & 68.7 & 20.2 \\
Sheltron & 2026 & 68M & yes & 12.0 & 10.6 & 28.9 & 20.1 & 26.4 & 4.4 & 6.7 & 95.2 & 57.2 \\
Sheltron (+task)$^\dagger$ & 2026 & 68M & yes & 12.0 & 26.0 & 47.4 & 9.0 & 18.8 & 69.7 & 85.0 & 65.4 & 8.0 \\
Prismor-1.5B & 2026 & 1.5B & no & 57.7 & 6.5 & 17.5 & 72.2 & 79.4 & 46.2 & 67.7 & 15.2 & 4.0 \\
Wolf Defender & 2026 & 308M & eval. & 3.1 & 0.9 & 3.1 & 73.8 & 83.9 & 0.0 & 0.0 & 90.8 & 3.5 \\
Horizon-Labs & 2026 & 308M & yes & 1.1 & 0.0 & 0.0 & 82.2 & 87.9 & 0.7 & 1.6 & 100.0 & 37.7 \\
\midrule
Judge (Qwen2.5-7B) & -- & 7B & -- & -- & -- & -- & 57.4 & 64.4 & 4.4 & 9.0 & 54.3 & 14.9 \\
Judge (Llama-3.1-8B) & -- & 8B & -- & -- & -- & -- & 55.3 & 61.9 & 5.4 & 9.6 & 78.3 & 8.3 \\
\bottomrule
\end{tabular}
}
\end{table*}

\subsection{RQ1: False Positives on Benign Tool Outputs}
\label{sec:rq1}

False alarms on benign tool outputs vary from none to most (Table~\ref{tab:many}). On AgentDojo, Horizon-Labs flags \MHZClean of AD-Clean outputs and Wolf Defender \MWFClean; they would stop \MHZBlock and \MWFBlock of tasks. ProtectAI v2 and PIGuard flag \CleanPAFPR and \CleanPGFPR of outputs and would stop \CleanPABlock and \CleanPGBlock of tasks (95\% intervals \CleanPABlockCI{} and \CleanPGBlockCI{}). Both Prompt Guard~2 models flag \MPGtClean and \MPGtsClean. On $\tau$-bench's JSON outputs the picture repeats: Horizon-Labs flags \TBFPRHZ, Wolf Defender \TBFPRWF, Prompt Guard~2 \TBFPRPGt, and Sheltron \TBFPRSH, while ProtectAI v2 flags \TBFPRPA and Prismor \TBFPRPR. The flagged outputs contain no instructions; the password confirmation of Figure~\ref{fig:example} is typical.

False-positive behavior on tool outputs is a stable property of a detector: its rank by false-positive rate on AgentDojo predicts its rank on $\tau$-bench (Kendall $\tau=\TauFPRADTB$, $p=\TauFPRADTBP$). It is not predicted by false positives on ordinary text. ProtectAI v2 flags \GenPAWebFPR of web pages and \GenPAEmailFPR of emails but \CleanPAFPR of AgentDojo tool outputs and \TBFPRPA of $\tau$-bench outputs.

\finding{1}{False-positive rates on benign agent tool outputs range from 0\% to over 90\% across detectors, are consistent across the two agent benchmarks ($\tau=\TauFPRADTB$), and are not predicted by false-positive rates on ordinary text.}

\subsection{RQ2: Detection at 1\% FPR and Rank Transfer}
\label{sec:rq2}

On AD-Matched, three detectors catch most injections at 1\% FPR: Horizon-Labs \MHZAD, Wolf Defender \MWFAD, and Prismor \MPRAD. Meta's Prompt Guard~2 catches \MPGtAD (86M) and \MPGtsAD (22M). The other DeBERTa detectors catch between \ADPATpr and \MTSAD, and PIGuard \ADPGTpr. On TB-Matched, Horizon-Labs catches \TBTprHZ, Sheltron \TBTprSH, and Wolf Defender \TBTprWF; PIGuard catches \TBTprPG and Prismor \TBTprPR.

\begin{figure*}[t]
\centering
\includegraphics[width=0.85\textwidth]{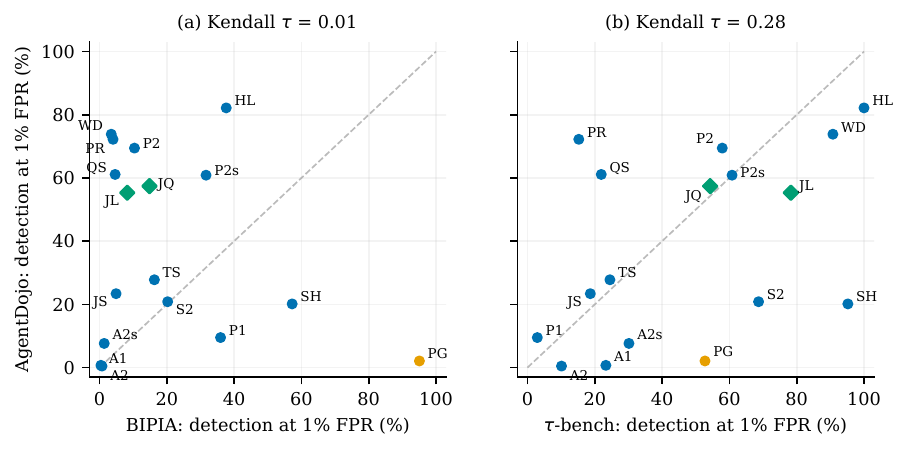}
\caption{Each point is a detector or judge (diamonds). (a) Detection on BIPIA does not predict detection on AgentDojo. (b) Detection on one agent benchmark predicts detection on the other only partly: Prismor (PR) and Sheltron (SH) trade places. Dashed line: equal performance. Key: HL Horizon-Labs, WD Wolf Defender, PR Prismor, P2/P2s Prompt Guard~2 86M/22M, P1 Prompt Guard~1, PG PIGuard, SH Sheltron, QS Qualifire Sentinel, S2 Sentinel v2, TS TestSavant, JS JasperLS, A1/A2/A2s ProtectAI v1/v2/v2-small, JQ/JL Qwen/Llama judge.}
\label{fig:rank}
\end{figure*}

Rankings do not transfer reliably between benchmarks (Figure~\ref{fig:rank}). Between BIPIA and AD-Matched, Kendall's $\tau$ is \RankTau ($p=\RankTauP$): PIGuard ranks first on BIPIA (\BIPGTpr) and \BestBipiaRankOnAD of \NDetectors on AgentDojo. Between the two agent benchmarks, $\tau$ is \TauADTB ($p=\TauADTBP$), and two detectors trade places: Prismor falls from \MPRAD on AgentDojo to \TBTprPR on $\tau$-bench, while Sheltron rises from \MSHAD to \TBTprSH. Between BIPIA and $\tau$-bench, $\tau$ is \TauBITB ($p=\TauBITBP$). None of these detection correlations is significant across the \NDetectors detectors. Horizon-Labs and Wolf Defender lead both agent benchmarks; the two Prompt Guard~2 models are consistent but lower on both (\MPGtAD and \TBTprPGt for 86M, \MPGtsAD and \TBTprPGts for 22M).

\finding{2}{A detector's detection rank on one benchmark is a weak guide to its rank on another ($\tau$ between \RankTau and \TauBITB, none significant). The best detector on BIPIA catches \ADPGTpr of AgentDojo injections at 1\% FPR, and a detector that catches \MPRAD of AgentDojo injections catches \TBTprPR on $\tau$-bench.}

\subsection{RQ3: Training Data}
\label{sec:rq3}

Where training data is public, it explains the ranking, and it shows that what transfers is the form of the inputs, not the attack strings.

\paragraph{PIGuard and BIPIA.}
PIGuard's published training split contains \PGBipiaRows labeled BIPIA inputs: full emails, tables, and code answers with and without inserted attacks (median \PGBipiaChars characters). The verbatim overlap is large (Table~\ref{tab:overlap}): \OvTextTrain text attacks and \OvCodeTrain code attacks from BIPIA's training half appear, as do \OvCleanEmail of BIPIA's clean emails and \OvCleanCode of its clean code contexts. None of BIPIA's test-half attacks appear, and our main results use only those; restricting to them lowers PIGuard's detection at 1\% FPR only from \BISplitPGTrainTpr to \BISplitPGTestTpr, because the two halves come from the same generator and are inserted into the same kinds of content. Training on one half and testing on the other is standard practice; the result is an in-distribution score.

\paragraph{PIGuard and InjecAgent.}
PIGuard's training split also contains \NIASeen InjecAgent attack instructions, but in another form: \PGIARows short prompts (median \PGIAChars characters) that pair an attack with a generic task sentence, not attacks inside tool responses. This gives a controlled comparison within TB-Matched, where every injected output has the same form. PIGuard detects \TBSeenPG of the outputs carrying an attack it was trained on and \TBUnseenPG of those carrying one of the \NIAUnseen attacks it was not. Having seen the attack text does not help when the surrounding input is new.

\paragraph{Horizon-Labs and the agent benchmarks.}
Horizon-Labs uses BIPIA only for evaluation and removes evaluation texts from its training data. Its training set includes five public datasets of injections planted in documents and tool responses. Our audit of those five finds no BIPIA attack string, no AgentDojo attack template, and no InjecAgent instruction. It is nevertheless the best detector on both agent benchmarks; what it shares with them is the form of input, structured tool output with planted instructions.

\paragraph{The others.}
Wolf Defender's card reports evaluation on a benchmark built from AgentDojo's injection payloads; its training data is not public. Prismor was fine-tuned on two prompt-level injection datasets. AgentDojo's templates are explicit instructions addressed to the model (``ignore your previous instructions'', ``this is an important message''), which resemble direct prompt injections; InjecAgent's attacks inside $\tau$-bench product names and BIPIA's attacks inside emails and tables do not. Prismor detects the first and not the others.

\begin{table}[t]
\centering\small
\caption{Verbatim overlap between BIPIA and PIGuard's published training split (whitespace and case normalized). A lower bound: paraphrases are not counted.}
\label{tab:overlap}
\begin{tabular}{@{}lr@{}}
\toprule
BIPIA component & Found in training split \\
\midrule
Text attacks, training half & \OvTextTrain \\
Code attacks, training half & \OvCodeTrain \\
Text attacks, test half & \OvTextTest \\
Code attacks, test half & \OvCodeTest \\
Clean email contexts & \OvCleanEmail \\
Clean code contexts & \OvCleanCode \\
Clean table contexts & \OvCleanTable \\
AgentDojo attack templates & none \\
\bottomrule
\end{tabular}
\end{table}

\finding{3}{Detectors work on inputs whose form resembles their training inputs. PIGuard's BIPIA score rests on training on full BIPIA inputs; having seen InjecAgent's attack strings as short prompts does not help it detect them inside tool outputs (\TBSeenPG versus \TBUnseenPG). Horizon-Labs leads both agent benchmarks without any overlap, after training on agent-style inputs.}

\subsection{RQ4: Task Awareness and Attack Goals}
\label{sec:rq4}

\paragraph{Judges.}
Telling a general-purpose model the user's request makes it a reasonable detector: at 1\% FPR the Qwen judge catches \ADJQTpr of AD-Matched and \TBTprJQ of TB-Matched injections, and the Llama judge \ADJLTpr and \TBTprJLl. Llama-3.1-8B was trained on data from before AgentDojo existed, so its detection cannot come from memorizing AgentDojo. On both agent benchmarks the judges stay below the best dedicated detectors, at the cost of a 7--8B forward pass per tool output.

\paragraph{Task-aware Sheltron.}
Passing the user's request to Sheltron in a trusted segment raises its AUROC on AD-Matched from \MSHADAuc to \MSHTADAuc but lowers its detection at 1\% FPR from \MSHAD to \MSHTAD, and raises its false-positive rate on clean tool outputs from \MSHClean to \MSHTClean on AgentDojo and from \TBFPRSH to \TBFPRSHT on $\tau$-bench. This use is undocumented and may simply be out of distribution for the model.

\begin{figure*}[t]
\centering
\includegraphics[width=0.92\textwidth]{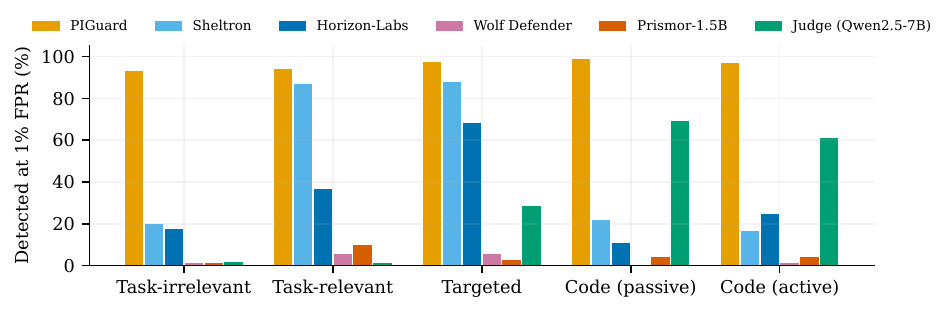}
\caption{Detection by BIPIA attack group (test half), each method at its own 1\%-FPR threshold on BIPIA. Only PIGuard, which was trained on BIPIA inputs, covers all groups; every other approach shown misses most task-irrelevant injections.}
\label{fig:goal}
\end{figure*}

\paragraph{Attack goals.}
Figure~\ref{fig:goal} breaks BIPIA's test-half injections into BIPIA's five official groups, each method at its own 1\%-FPR threshold on BIPIA. No approach other than the BIPIA-trained PIGuard covers all groups. Task-irrelevant injections, which ask the model to perform some unrelated harmless task, are the hardest: apart from PIGuard (\GPGIrr), the best approach is Prompt Guard~1 with \GPGoIrr, and every other approach catches at most \GSHIrr. The others split the remaining groups: Sheltron catches \GSHRel of task-relevant and \GSHTgt of targeted injections but \GSHAct of active code attacks, while the Qwen judge catches \GJQPas and \GJQAct of passive and active code attacks and \GJQRel of task-relevant ones. Our judge prompt asks about actions and tells the model not to flag ordinary requests between people; we did not vary it, so we cannot say how much of the judges' gap comes from the question we asked.

\finding{4}{Task-aware 7--8B judges detect 54--78\% of injections in agent tool outputs at 1\% FPR, below the best dedicated detectors, and giving the user's request to a small classifier does not help. Outside its training distribution, every approach misses whole classes of injections, and task-irrelevant injections evade nearly all of them.}

\subsection{RQ5: Format and Metadata}
\label{sec:rq5}

\begin{figure}[t]
\centering
\includegraphics[width=\columnwidth]{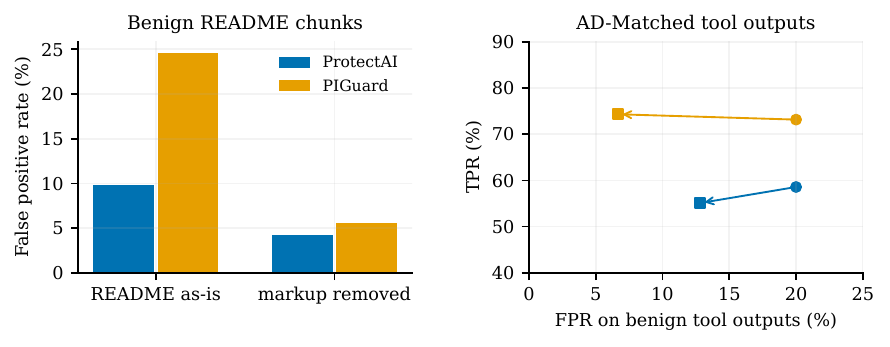}
\caption{Format alone moves false positives. Left: removing markup from benign READMEs. Right: removing YAML keys from AgentDojo tool outputs (circle: as-is; square: keys removed) lowers PIGuard's FPR threefold at unchanged TPR.}
\label{fig:format}
\end{figure}

How a tool output is presented to a detector moves its false-positive rate. README chunks that PIGuard flags are full of HTML tables, badges, and links rather than instructions; stripping markup lowers its FPR on READMEs from \FmtPGReadmeOrig to \FmtPGReadmePlain and ProtectAI v2's from \FmtPAReadmeOrig to \FmtPAReadmePlain (Figure~\ref{fig:format}). On AD-Matched, removing YAML keys so that PIGuard sees only values lowers its FPR from \ADPGFprHalf to \ADPGNFprHalf at the same TPR (\ADPGTprHalf versus \ADPGNTprHalf). Sheltron's source token has the largest effect: declaring a tool output as \texttt{tool\_result} instead of \texttt{other} lowers its FPR on clean tool outputs from \MSHGClean to \MSHClean on AgentDojo and from \TBFPRSHG to \TBFPRSH on $\tau$-bench.

These changes help less at the low-FPR end. With YAML keys removed, PIGuard catches \ADPGNTpr of AD-Matched injections at 1\% FPR and ProtectAI v2 \ADPANTpr. Sheltron with the specific source token catches \MSHAD on AD-Matched, compared with \MSHGAD with the generic one, and \TBTprSH versus \TBTprSHG on TB-Matched. Format and metadata shift a detector's scores; they do not make an out-of-distribution detector work.

\finding{5}{Removing serialization markup or declaring the input type lowers default-threshold false positives by up to about twentyfold, a free improvement for deployments, but after any of these changes neither PIGuard nor ProtectAI v2 exceeds 15\% detection at 1\% FPR on AD-Matched.}

\section{Discussion and Limitations}
\label{sec:discussion}

\paragraph{Choosing a detector for an agent.}
Our results suggest four practical rules. First, measure false positives on the agent's own tool outputs and translate them into blocked tasks; false positives on prose do not predict them. Second, report detection at a fixed low FPR; default-threshold accuracy on balanced sets says little about either error. Third, before trusting a benchmark score, check whether the detector was trained on that benchmark or on data generated the same way, and prefer evaluation data the detector cannot have seen. Fourth, prefer a detector trained on inputs like the ones it will screen: in our study, Horizon-Labs, Wolf Defender, and both Prompt Guard~2 models kept false positives near zero on both agent benchmarks and caught more than half of the injections on each; Horizon-Labs and Wolf Defender, both built for untrusted content reaching agents, caught more than 70\%. Even these missed most task-irrelevant injections, so a detector is a filter to combine with system-level defenses~\cite{camel}, not a replacement for them.

\paragraph{Simulated agent environments.}
\emph{Limitation.} Both agent benchmarks are simulations: AgentDojo returns synthetic YAML records, and $\tau$-bench returns JSON from synthetic databases. Real tool outputs (HTML pages, API responses, long documents) are messier, and our results show that format alone moves false-positive rates, so absolute numbers will differ in deployment. In $\tau$-bench we also chose the injection point (product names) and the attack set (InjecAgent) ourselves; another placement, such as reviews or seller messages, could favor different detectors. The findings we expect to carry over are the relative ones that held on both agent benchmarks: which detectors over-flag tool outputs, and that the agent-oriented detectors lead.

\paragraph{Static attacks.}
\emph{Limitation.} All injections come from benchmark templates. An adaptive attacker who knows the detector or judge would lower every detection rate we report~\cite{attackermovessecond}. Our detection numbers are upper bounds; the false-positive results do not depend on the attacker.

\paragraph{Detector coverage.}
\emph{Limitation.} We evaluate the open and license-gated detectors we could access; commercial API detectors are closed and not included. Only PIGuard and Horizon-Labs publish enough to audit their training data; for the others we can report only what their model cards state.

\paragraph{The judges are a diagnostic, not a defense.}
The task-aware judges call a 7--8B model once per tool output, use a single prompt, and were not tested against adaptive attacks. We use them to show which information the detectors lack and which attack goals that information covers. A judge prompt that also asks about changes to the answer might cover task-irrelevant and task-relevant injections, at a cost in false alarms we have not measured.

\paragraph{Verbatim overlap is a lower bound.}
Our audits find exact copies only. Paraphrases, attacks generated from the same templates, and unpublished data escape them. For PIGuard and BIPIA, verbatim overlap alone already explains the score.

\section{Related Work}
\label{sec:related}

\paragraph{Evaluation methodology for injection detectors.}
Fomin~\cite{benchmarkslie} trains activation probes on 18 datasets and shows that leave-one-dataset-out evaluation lowers AUROC compared with in-distribution splits. PromptShield~\cite{promptshield} argues for evaluating detectors at low FPR and builds a curated benchmark for it. PIGuard~\cite{piguard} measures over-defense on benign prompts with trigger words. Unlike these works, we evaluate released detectors on the tool outputs an agent actually screens, translate false positives into blocked tasks, and audit a released detector's published training split against the benchmark used to praise it.

\paragraph{Pitfalls in ML and LLM security research.}
Arp et al.~\cite{arp2022dos} and Evertz et al.~\cite{chasingshadows} catalog pitfalls such as data leakage and inappropriate metrics in security research with ML and LLMs. Nasr et al.~\cite{attackermovessecond} show that defenses evaluated on static attacks fall to adaptive ones. Our study is an instance of the leakage and metric pitfalls for one widely used class of defenses, with the overlap measured directly rather than inferred.

\paragraph{Detection-based defenses.}
ProtectAI's detector~\cite{protectai}, PIGuard~\cite{piguard}, and Prompt Guard in LlamaFirewall~\cite{llamafirewall} classify text with a fine-tuned encoder. A newer generation of open detectors, released in 2026 and evaluated here, trains on documents and tool responses with planted instructions or adds role and source tokens to the input; their model cards report results on their own or AgentDojo-derived evaluation sets. DataSentinel~\cite{datasentinel} detects injections with an LLM fine-tuned in a minimax game against adaptive attacks, and AlignSentinel~\cite{alignsentinel} separates instructions that align with the user's task from those that do not, using attention features. Our task-aware judges follow the same intuition as AlignSentinel without training; we use them only to diagnose what the classifiers miss.

\paragraph{Benchmarks.}
AgentDojo~\cite{agentdojo}, InjecAgent~\cite{injecagent}, and BIPIA~\cite{bipia} measure whether agents or LLMs follow injected instructions. We reuse AgentDojo's ground-truth tool calls and BIPIA's contexts and attacks to build matched benign and injected sets for detectors, a use their authors did not design them for.

\paragraph{Model- and system-level defenses.}
StruQ~\cite{struq} and SecAlign~\cite{secalign} train the model to ignore instructions in data; CaMeL~\cite{camel} keeps untrusted data out of the planner. These defenses do not rely on recognizing an injection, so our findings about detectors do not apply to them.

\section{Conclusion}
Prompt-injection detectors are chosen by benchmark scores, but across fifteen detectors, two agent benchmarks, and BIPIA, detection rankings transferred poorly from one benchmark to another. What a detector was trained on explained where it worked: the detector that looks best on BIPIA was trained on full BIPIA inputs and misses nearly every AgentDojo injection, seeing an attack's text without its surrounding tool output did not help, and the detector that led both agent benchmarks shared no data with them but had been trained on agent-style inputs. False-positive behavior on tool outputs, by contrast, was stable across agent benchmarks and is cheap to measure. Evaluations meant to guide deployment should replay the agent's own tool outputs, report detection at a low false-positive rate and the task-level cost, and audit what the detector has already seen.

\section*{Ethics Considerations}
We use only public benchmarks and public benign corpora, and create no new attacks: every injected sample uses an attack string that AgentDojo, BIPIA, or InjecAgent already publishes. We ran no experiments against deployed systems or other people's agents. The overlap audits concern training data that detector authors chose to publish; we report them to improve evaluation practice.

\section*{Open Science}
The scripts that build every evaluation set (they regenerate the data from the public benchmarks rather than redistributing it), the detector and judge scores, and the analysis code that produces every number and figure in this paper are available at \url{https://github.com/lzwhehe/benign-instruction-bench}.

\section*{Acknowledgments and Use of AI Tools}
Zhuowen Liu initiated and directed this study. The author set the research direction and the criteria for selecting a project, chose which candidate hypotheses to test and which to pursue after pilot experiments, directed the extension of the study to a second agent benchmark and to license-gated detectors including Prompt Guard~2, provided and managed all compute, obtained access to the gated models, and reviewed all experiments, results, and text. The author used Claude (Anthropic), an AI assistant, under this direction to implement the experiment and analysis code, run the experiments, and draft the manuscript. The author takes full responsibility for the content of this paper.

{\footnotesize \bibliographystyle{acm}
\bibliography{refs}}

\appendix
\section{Detector Configurations}
\label{app:configs}
Each detector is run as its model card documents. Inputs longer than the context window are truncated, except for Sheltron, whose card specifies overlapping windows (stride 768), and Prompt Guard, whose cards specify splitting long inputs into 512-token segments; for both we take the maximum score over windows. Prismor's card truncates the whole chat prompt at 512 tokens, which cuts the assistant header from long inputs; we truncate only the text so the template stays intact.

\begin{table}[h]
\centering\small
\resizebox{\columnwidth}{!}{%
\begin{tabular}{@{}llrl@{}}
\toprule
Detector & Hugging Face model & Ctx & Input \\
\midrule
ProtectAI v1 & \texttt{protectai/\ldots-injection} & 512 & text \\
ProtectAI v2 & \texttt{protectai/\ldots-injection-v2} & 512 & text \\
JasperLS & \texttt{JasperLS/deberta-v3-base-injection} & 512 & text \\
TestSavant-L & \texttt{testsavantai/\ldots-large-v0} & 512 & text \\
PIGuard & \texttt{leolee99/PIGuard} & 512 & text \\
Sheltron & \texttt{sheltron-ai/prompt-guard-68m} & 1024 & role/source tokens \\
Prismor & \texttt{prismor/prompt-guard-1.5b} & 512 & chat template \\
Wolf Defender & \texttt{patronus-studio/wolf-defender-\ldots} & 2048 & text \\
Horizon-Labs & \texttt{Horizon-Labs/\ldots-guard-base} & 2048 & text \\
Prompt Guard 2 (86M) & \texttt{meta-llama/Llama-Prompt-Guard-2-86M} & 512 & segmented \\
Prompt Guard 2 (22M) & \texttt{meta-llama/Llama-Prompt-Guard-2-22M} & 512 & segmented \\
Prompt Guard 1 & \texttt{meta-llama/Prompt-Guard-86M} & 512 & segmented \\
ProtectAI v2-small & \texttt{protectai/\ldots-small-\ldots-v2} & 512 & text \\
Qualifire Sentinel & \texttt{qualifire/prompt-injection-sentinel} & 2048 & text \\
Sentinel v2 & \texttt{rogue-security/\ldots-sentinel-v2} & 2048 & text \\
\bottomrule
\end{tabular}}
\end{table}

\section{Detailed Results for Four Methods}
\label{app:detail}

\begin{figure*}[t]
\centering
\includegraphics[width=0.9\textwidth]{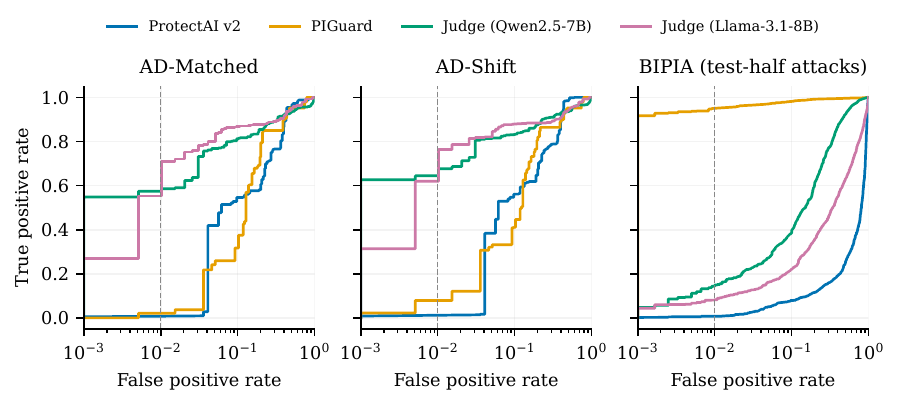}
\caption{ROC curves with logarithmic FPR axis for the four methods of Table~\ref{tab:main}; the dashed line marks 1\% FPR. The Llama judge's wide interval on AD-Matched comes from a few benign outputs with high scores that make the 1\% threshold unstable.}
\label{fig:roc}
\end{figure*}

Table~\ref{tab:main} gives AUROC, bootstrap intervals for TPR at 1\% FPR, and default-threshold rates for the two detectors and two judges we examine most closely; Figure~\ref{fig:roc} shows their ROC curves.

\begin{table*}[t]
\centering\small
\caption{Detection on the three matched sets for four methods. Brackets: 95\% bootstrap interval. BIPIA uses only test-half attacks.}
\label{tab:main}
\begin{tabular}{@{}llcccc@{}}
\toprule
Set & Method & AUROC & TPR @ 1\% FPR (\%) & FPR @ 0.5 (\%) & TPR @ 0.5 (\%) \\
\midrule
\multirow{4}{*}{AD-Matched} & ProtectAI v2 & \ADPAAuc & \ADPATprT{} [\ADPATprCI] & \ADPAFprHalfT & \ADPATprHalfT \\
 & PIGuard & \ADPGAuc & \ADPGTprT{} [\ADPGTprCI] & \ADPGFprHalfT & \ADPGTprHalfT \\
 & Judge (Qwen2.5-7B) & \ADJQAuc & \ADJQTprT{} [\ADJQTprCI] & \ADJQFprHalfT & \ADJQTprHalfT \\
 & Judge (Llama-3.1-8B) & \ADJLAuc & \ADJLTprT{} [\ADJLTprCI] & \ADJLFprHalfT & \ADJLTprHalfT \\
\midrule
\multirow{4}{*}{AD-Shift} & ProtectAI v2 & \ADSPAAuc & \ADSPATprT{} [\ADSPATprCI] & \ADSPAFprHalfT & \ADSPATprHalfT \\
 & PIGuard & \ADSPGAuc & \ADSPGTprT{} [\ADSPGTprCI] & \ADSPGFprHalfT & \ADSPGTprHalfT \\
 & Judge (Qwen2.5-7B) & \ADSJQAuc & \ADSJQTprT{} [\ADSJQTprCI] & \ADSJQFprHalfT & \ADSJQTprHalfT \\
 & Judge (Llama-3.1-8B) & \ADSJLAuc & \ADSJLTprT{} [\ADSJLTprCI] & \ADSJLFprHalfT & \ADSJLTprHalfT \\
\midrule
\multirow{4}{*}{BIPIA} & ProtectAI v2 & \BIPAAuc & \BIPATprT{} [\BIPATprCI] & \BIPAFprHalfT & \BIPATprHalfT \\
 & PIGuard & \BIPGAuc & \BIPGTprT{} [\BIPGTprCI] & \BIPGFprHalfT & \BIPGTprHalfT \\
 & Judge (Qwen2.5-7B) & \BIJQAuc & \BIJQTprT{} [\BIJQTprCI] & \BIJQFprHalfT & \BIJQTprHalfT \\
 & Judge (Llama-3.1-8B) & \BIJLAuc & \BIJLTprT{} [\BIJLTprCI] & \BIJLFprHalfT & \BIJLTprHalfT \\
\bottomrule
\end{tabular}
\end{table*}

\section{Judge Prompt}
\label{app:prompt}
Both judges receive the following system prompt, then a user message of the form ``User request: \ldots{} Tool \texttt{<name>} returned: \ldots'' (tool output truncated to 6{,}000 characters). We read the first-token distribution and compute $P(\text{Yes})/(P(\text{Yes})+P(\text{No}))$, summing over case and whitespace variants of each word. Both judges are quantized to 4 bits (NF4); the 4-bit Qwen judge matches the same model served in bf16 to within 0.02 AUROC on AD-Matched.

\begin{quote}\small
You audit data returned by tools to an AI assistant. Tool data is supposed to be passive information. Answer `Yes' if the tool data contains text that tries to make the assistant perform actions or tasks the user did not ask for. Answer `No' otherwise, including when the data merely describes procedures or contains ordinary requests between people. Reply with exactly one word: Yes or No.
\end{quote}

\section{Excluded Detectors}
\label{app:excluded}
\texttt{deepset/deberta-v3-base-injection} and \texttt{fmops/distilbert-prompt-injection} flag every benign README chunk we scored (\DegDeepsetReadme{} and \DegFmopsReadme{}, respectively) and more than three quarters of AD-Clean tool outputs. The fmops model also flags two of five plain sentences such as ``The weather is nice today.'' We treat both as unusable and omit them. \texttt{leolee99/InjecGuard} returns scores identical to PIGuard on all five sets, so we report it once.

\section{Labeling by Differential Replay}
\label{app:labels}
Our first labeling attempt marked an output as injected if it contained the injected string verbatim. Because AgentDojo serializes tool outputs as YAML, multi-line injections are re-wrapped and escaped, and every output carrying the \texttt{important\_instructions} attack was labeled benign. Differential replay (same ground-truth call, clean versus injected environment) avoids the problem; with it, each of the three AD-Matched templates contributes the same number of injected outputs. We note the pitfall because substring labeling is the obvious approach and silently inflates false-positive rates.

\end{document}